\documentclass{article}
\pdfoutput=1
\usepackage{spconf,amsmath,graphicx}
\usepackage{xcolor}
\usepackage{tabularx}
\usepackage{hyperref}

\definecolor{myblue}{RGB}{108, 142, 191}
\definecolor{mygreen}{RGB}{130, 179, 102}

\title{COMPARISON OF SPEAKER ROLE RECOGNITION AND SPEAKER ENROLLMENT PROTOCOL FOR CONVERSATIONAL CLINICAL INTERVIEWS}
\name{ \begin{tabular}{c} Rachid Riad$^{\star 1,2}$ \thanks{$\star$ Equal contribution.}, Hadrien Titeux$^{\star  1}$, Laurie Lemoine$^{2,3}$, Justine Montillot$^{2,3}$, \\Agnes Sliwinski$^{2,3}$, Jennifer Hamet Bagnou$^{2,3}$, Xuan Nga Cao$^1$,\\ Anne-Catherine Bachoud-Lévi$^{2,3}$, Emmanuel Dupoux$^{1}$\end{tabular}}
\address{
  $^1$ CoML/ENS/CNRS/EHESS/INRIA/PSL Research University, Paris, France \\
  $^2$ NPI/ENS/INSERM/UPEC/PSL Research University, Cr\'eteil, France  \\
    $^3$ Huntington's Disease National Reference Center, Neurology Department, \\
   Henri-Mondor Hospital, APHP, Cr\'eteil, France
  }
\begin{document}
%
\maketitle
\begin{abstract}
Conversations between a clinician and a patient, in natural conditions, are valuable sources of information for medical follow-up. The automatic analysis of these dialogues could help extract new language markers and speed-up the clinicians' reports. Yet, it is not clear which speech processing pipeline is the most performing to detect and identify the speaker turns, especially for individuals with speech and language disorders. Here, we proposed a split of the data that allows conducting a comparative evaluation of speaker role recognition and speaker enrollment methods to solve this task. We trained end-to-end neural network architectures to adapt to each task and evaluate each approach under the same metric. Experimental results are reported on naturalistic clinical conversations between Neuropsychologist and Interviewees, at different stages of Huntington’s disease.
We found that our Speaker Role Recognition model gave the best performances. In addition, our study underlined the importance of retraining models with in-domain data. Finally, we observed that results do not depend on the demographics of the Interviewee, highlighting the clinical relevance of our methods.
\end{abstract}
 \vspace{-0.5em}
\section{Introduction}
\label{sec:intro}
 \vspace{-0.5em}

 During the last decades, it became easier to collect large naturalistic corpora of speech data. It is now possible to obtain new realistic measurements of turn-takings and linguistic behaviors \cite{ash2015study}. These measurements can be especially useful during clinical interviews as they augment the current clinical panel of assessments and unlock  home-based assessments \cite{matton2019into}.
The remote automatic measure of symptoms of patients with Neurodegenerative diseases could greatly improve the follow-up of patients and speed-up ongoing clinical trials.

Yet, this methodology relies on the heavy burden of manual annotation to reach the necessary amount needed to draw significant conclusions. It is now indispensable to have robust speech processing pipelines to extract meaningful insights from these long naturalistic datasets. Huntington's Disease represents a unique opportunity to design and test these speech algorithms for \textit{Neurodegenerative diseases}. Indeed, individuals with the Huntington's disease can exhibit a large spectrum of \textit{speech and language} symptoms \cite{vogel2012speech} and it is possible to follow gene carriers even before the official clinical onset of the disease \cite{hinzen2018systematic}. The first unavoidable computational tasks to extract speech and linguistic information from medical interviews are: (1) the \textit{detection} of speaker-homogeneous portions of voice activity \cite{graf2015features} and (2) the \textit{identification} of the speaker \cite{bigot2010looking}. Speaker Turns are clinically informative for the diagnostic in Huntington's Disease \cite{perez2018classification, vogel2012speech}.

\begin{figure}[!t]
    \centering
    \caption{Two approaches for the detection and identification of segments for conversational clinical interviews. The steps for the Speaker Enrollment Protocol are in \textcolor{myblue}{Blue}, and \textcolor{mygreen}{Green} for the Speaker Role Recognition.}
    \includegraphics[width=0.33\textwidth]{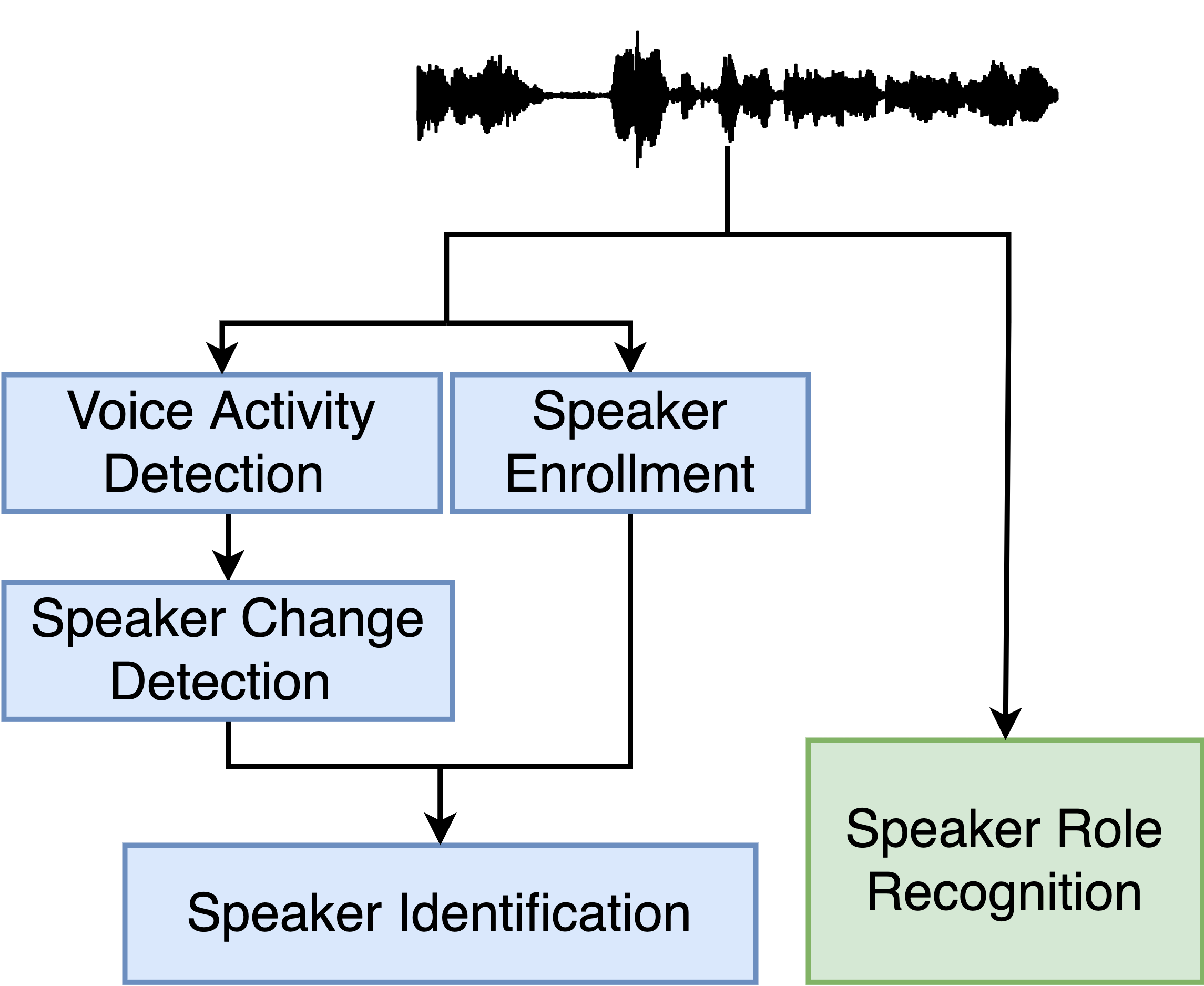}
\vspace{-1.em}
    \label{fig:pipeline}
\end{figure}

First, a number of studies are trying to solve this problem directly from the audio signal and linguistic outputs, also referred to as \textit{Speaker Role Recognition}. They are taking advantage of the specificities (ex: prosody, specific vocabulary, adapted language models) of each role in the different domains: Broadcast news programs \cite{bigot2010looking}, Meetings \cite{sapru2012automatic}, Medical conversations \cite{flemotomos2018combined}, Child-centered recordings \cite{lavechin2020open}.

Another approach relies on \textit{Speaker Enrollment} \cite{snyder2017deep, heigold2016end}, it aims to check the identity of a given speech segment based on a enrolled speaker template. Our study differ from these studies as they are evaluating their pipelines with already segmented speaker-homogeneous speech segments.
Another related approach is the \textit{Personal VAD} (Voice Activity Detection) model from \cite{ding2020personal} where they used an enrolled speaker template to detect speech segments from each individual speaker, on the Librispeech Dataset.

None of these approaches have been compared under the same evaluation metric, despite prior works aiming at solving both these tasks \cite{garcia2019speaker} and their high degree of similarities.

Here in this paper, we aimed to \textit{detect} automatically the portions of speech and to \textit{identify} the speakers in medical conversation between Neuropsychologists and Interviewees. These interviewees are either Healthy Controls (C), gene carriers without overt manifestation of Huntington's Disease (preHD) and manifest gene carriers of Huntington's Disease (HD). We used two different speech processing approaches to deal with these 2 problems (Figure \ref{fig:pipeline}): \textit{Speaker Role Recognition} and  \textit{Speaker Enrollment Protocol}.
\vspace{-1.em}
\section{Data, evaluation splits, metrics}
\vspace{-0.5em}
\subsection{Dataset}
\vspace{-0.5em}
Ninety four participants were included from two observational cohorts (NCT01412125 and NCT03119246) in this ancillary study at the Hospital Henri-Mondor Créteil, France): 72 people tested  with a number of CAG repeats on the Huntingtin gene above 35 \cite{gusella1983polymorphic} (CAG $>35$), and 22 Healthy Controls (C). Mutant Huntington gene carriers were considered premanifest if they both score less than five at the Total Motor score (TMS) and their Total functional capacity (TFC) equals 13 \cite{tabrizi2009biological} using the Unified Huntington Disease Rating Scale (UHDRS) \cite{kieburtz2001unified}. The demographics are summarized in Table \ref{tab:demographic_patients}. All participants signed an informed consent and conducted an interview with an expert neuropsychologist. Therefore in the current setting, there are two roles in each conversational interview: a \textit{Neuropsychologist} and an \textit{Interviewee}.
The speech data were annotated with Seshat \cite{seshat2020titeuxriad} and Praat \cite{boersma2002praat} softwares. The annotators were second-year graduate students in speech pathology, all French native speakers. The recordings were done in the same condition for all participants, with a ZOOM H4n Pro recorder, sampled at 44.1 kHz with a 16-bit resolution.
\vspace{-0.5em}

\begin{table}[!ht]
\vspace{-1em}
\caption{Interviewee demographics and clinical scores}
	\centering

	 \begin{tabular}{|p{0.24\linewidth}|c|c|c|}

		\hline
		 & Controls & \multicolumn{2}{c|}{Huntington's disease } \\
		 &  & \multicolumn{2}{c|}{Gene carriers} \\
		 \hline
		                Sub-groups            & C  & PreHD     &  HD          \\
		\hline

		N               & 22  & 18 & 54       \\

		         Gender        & 10F/12M   & 10F/8M  & 32F/22M    \\
				Age  (years)                 &  54.1 (8.6)  & 50.1 (11.8) & 53.5 (11.3)      \\
				 CAG Triplets                   & $\leq$ 35  & 41.5 (1.7) & 44.2 (3.3)       \\
				 \hline
				 TFC     \cite{shoulson1981huntington}             & ---  & 13.0 (0.0) & 10.4 (2.1)       \\
					TMS               & ---  & 0.33 (1.0) & 34.3 (15.6)       \\
		\hline
	\end{tabular}
	    \label{tab:demographic_patients}
\vspace{-1em}
\end{table}

\vspace{-0.5em}
\subsection{Splits of the data}
\vspace{-0.5em}

\begin{figure}[!ht]
    \centering
    \vspace{-1.em}
    \caption{Illustration of the data split with 4 interviews. Each line $I_i$ represents an interview between the Interviewee and the Neuropsychologist. The elevation of each row indicates 'who speaks when'. The segments can overlap.     }
    \includegraphics[width=0.48\textwidth]{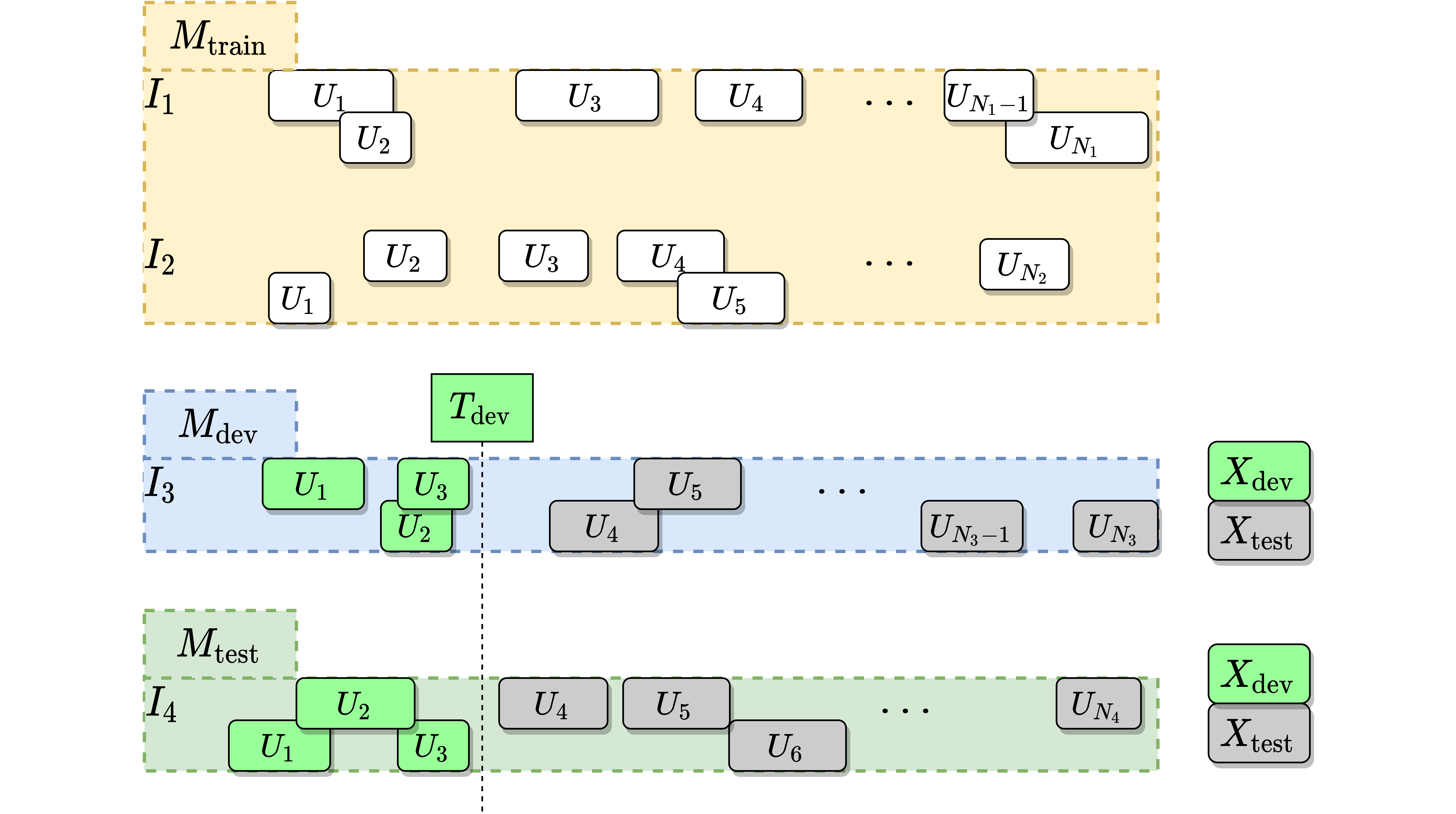}
\vspace{-2.em}
    \label{fig:split_data}
\end{figure}

The dataset is composed of $K=94$ interviews $\mathcal{I}_{1\ldots K}$. We designed a split of the dataset to compare the Speaker Role Recognition and Speaker Enrollment pipelines (See Figure \ref{fig:split_data}).
The dataset is split in three sets which we refer to \textit{meta-train set} $M_{train}$, \textit{meta-dev set} $M_{dev}$ and \textit{meta-test set} $M_{test}$  with the ratio of 60\%, 20\%, and 20\%, respectively. Interview $I \in \mathcal{I}_{1\ldots K}$ is composed of $N_I$ segments $I=\left\{U_{0}, U_{2}, \ldots, U_{N_I}\right\}$.
Each segment $U_i$ is pronounced by a speaker $s_i$. We summarized the corpus statistics in the Table \ref{tab:corpus_split_stats}.

Each interview $I$ in  the \textit{meta-dev} and \textit{meta-test} is split in two sets which we refer \textit{dev set} $X_{dev}$ and \textit{test set} $X_{test}$. $X_{test}$ is always kept fixed through all experiments, and we study the influence of the size of the $X_{dev}$ based on $T_{\text{dev}}$ that filters the segments (cf Figure \ref{fig:split_data}).

All the data from the \textit{meta-train} set $M_{train}$ is used to train or fine-tune the neural network models for voice activity detection, speaker change detection, speaker role recognition, and speaker enrollment. The dev set $X_{dev}$ of the \textit{meta-dev} set $M_{dev}$ and the dev set $X_{dev}$ of the \textit{meta-test} set $M_{test}$ are only used for the speaker enrollment experiments, to build the template representation of each speakers. The results on the test set $X_{test}$ of the \textit{meta-dev} set $M_{dev}$ are used to select all the hyper-parameters and select the best model for each experiment. The final comparison is done with the test set $X_{test}$ of the \textit{meta-test} set $M_{test}$.

\vspace{-0.5em}

\begin{table}[!ht]
    \centering
    \caption{Corpus statistics. NP stands for Neuropsychologist. IT stands for Interviewee. \textit{Dur} stands for Duration and reported in hour. }
\begin{tabular}{|l|c|c|c|}
\hline
     &  $M_{train}$& $M_{dev}$& $M_{test}$ \\
     \hline
    \#Interviews &  $57$& $18$& $19$ \\
        \#Segments IT&  $21400$& $7503$& $7788$\\
        \#Segments NP&  $4184$& $1381$& $1517$\\
        \textit{Dur} Role IT (h) &  $7.65$& $3.02$& $3.21$ \\
        \textit{Dur} Role NP (h)&  $ 3.54 $& $1.14$& $1.15$ \\
        \textit{Dur} Overlap (h)&  $1.10$& $0.50$& $0.45$ \\
         \#(C/preHD/HD) &  $13/11/33$& $4/4/10$& $5/3/11$ \\
    \hline

\end{tabular}
\vspace{-1.em}
    \label{tab:corpus_split_stats}
\end{table}

\vspace{-1.em}
\subsection{Metrics}
\vspace{-0.5em}
\label{subsec:overall_metrics}

To compare the final performance of the pipeline systems, we use the Identification Error Rate (IER) taking into account both the segmentation errors and confusion errors. We obtained the IER with {\small \texttt{pyannote.metrics}} \cite{pyannote.metrics}.
\begin{equation*}
    	\text{IER} = \frac{T_\text{false alarm} + T_\text{missed detection} + T_\text{confusion}}{T_\text{Total}}
\end{equation*}
The $\frac{T_\text{confusion}}{T_\text{Total}}$ component in the IER is related to the Miss-classification Rate (MR\%) used in Speaker Role Recognition study \cite{flemotomos2019role}, which is based on Frames and not duration of the turns. We compared the different approaches as a function of the size of the enrollment $T_{dev}$ in Figure \ref{fig:results_t_dev}. Regarding the relevance for Healthcare applications, we showed the IER details in Figure \ref{fig:results_per_class}, for both best approaches based on the Interviewee demographics.
\vspace{-1.5em}
\section{Methods}
\vspace{-0.5em}
\subsection{Speaker Role Recognition}
\vspace{-0.5em}
We used a modified approach from \cite{lavechin2020open} for the Speaker Role Recognition. We trained on $M_{train}$ a unique model to detect each role (Neuropsychologist,Interviewee), and selects the best epoch on $M_{dev}$.  This is a multi-label multi-class segmentation problem. A threshold parameter for each role is optimized on the Meta-dev set $M_{dev}$ for the two output units of the model.
Therefore the two classes can be activated at the same time, i.e. we can also detect overlapped speech.

To solve and model this task, we used SincNet filters \cite{ravanelli2018speaker} to obtain adapted speech features vectors from the audio signal. The SincNet output is fed to a stack of 2 bi-recurrent LSTM layers with hidden size of 128, then pass to a stack of 2 feed-forward layers of size 128 before a final decision layer. We used a binary cross-entropy loss and a cyclic scheduler as training procedure. The version of this model and the hyper-parameters used to train it, can be found  (\href{https://github.com/MarvinLvn/voice-type-classifier/blob/new_model/model/config.yml}{here}).
\vspace{-1.em}
\subsection{Speaker enrollment protocol}
\vspace{-0.5em}
\label{sec:spk_enrollment}
The Speaker enrollment protocol can be decomposed into four tasks: (1) Voice Activity Detection (2) Speaker Change Detection,  (3) Enrollment, (4) Identification. We extended the speech processing toolkit from \cite{bredin2020pyannote} {\small \texttt{pyannote.audio}} to run our experiments. Clinical laboratories can not all re-train in-domain speech processing models due to data scarcity or a lack of computational resources. Therefore, we evaluated pretrained models on open-source datasets and transfer models on our dataset to evaluate these out-of-domain performances with real clinical conversational conditions.
\vspace{-1.em}
\subsubsection{Voice Activity Detection}
\vspace{-0.5em}
The first step is the Voice Activity Detection (VAD), i.e. obtain the speech segments in the audio signal. It can be modeled as an audio sequence labeling task. There are $2$ classes (Speech or Non-Speech).
The VAD labels for each interview I are the presence or not of a segment $U_i$ at time $t$.

The model can be used already \textit{Pretrained} or \textit{Retrained} on the meta-train set $M_{train}$ of our dataset. We choose the DIHARD dataset \cite{ryant2019second} as a potential pretrained dataset as it contains multiple source domain data (clinical interviews among them). When trained from scratch, the training is done for 200 pyannote epochs and the model is selected on the Meta-dev $M_{dev}$. The model is also composed of SincNet filters with 2 bi-recurrent LSTM layers and 2 feed-forward layers. The full specifications can be found \href{https://github.com/pyannote/pyannote-audio/tree/65a50951295b85fa589cac75c6673fb979a7858b/tutorials/models/speech_activity_detection}{here}.
\vspace{-1em}
\subsubsection{Speaker Change Detection}
\vspace{-0.5em}
The second step is the Speaker Change Detection (SCD), i.e. obtain the moment when one of a speaker starts or stops talking. It can aslo be modeled as an audio sequence labeling task. There are $2$ classes (Change or No-Change). The SCD labels for each interview I are the start or end of a segment $U_i$ at time $t$. We also compared \textit{Pretrained} on DIHARD and \textit{Retrained} models. We used the same model as for the Voice Activity Detection. The full specifications can be found \href{https://github.com/pyannote/pyannote-audio/tree/65a50951295b85fa589cac75c6673fb979a7858b/tutorials/models/speaker_change_detection}{here}.

Based on VAD and SCD outputs, we can now obtain for each Interview $I$ a set of $N'_I$ candidates  speaker-homogeneous segments $\{\hat{U}_1, \dots \hat{U}_{N'_I}\}$.
\vspace{-1.em}
\subsubsection{Enrollment}
 \vspace{-0.5em}
In the enrollment stage, we need to get a Speaker Embedding function $f_\theta$ for our specific task. We combined SincNet filters and the X-vector architecture \cite{snyder2017deep} as in \cite{bredin2020pyannote}.

 The model can be used already \textit{Pretrained} or \textit{Finetuned} on the meta-train set $M_{train}$ of our dataset \footnote{Our dataset is too small to train from scratch the X-vector model.}. For finetuning, we froze all layers and fine-tuned the last layer. We used the VoxCeleb2 dataset \cite{nagrani2017voxceleb} as a pretraining dataset as it contains a diverse distribution of speakers and recording conditions.
 \begin{figure}[!ht]
    \centering
    \caption{Identification Error Rates for the different combination of approaches on the test set $X_{test}$ of the meta-test set $M_{test}$ as a function of the size of the enrollment parameter $T_{dev}$. \textit{Spk Emb.}, \textit{VAD},\textit{SCD} stand for Speaker Embedding, Voice Activity Detection and Speaker Change Detection. Best performance of each approach displayed at the best $T_{dev}$.}
    \includegraphics[width=0.475\textwidth]{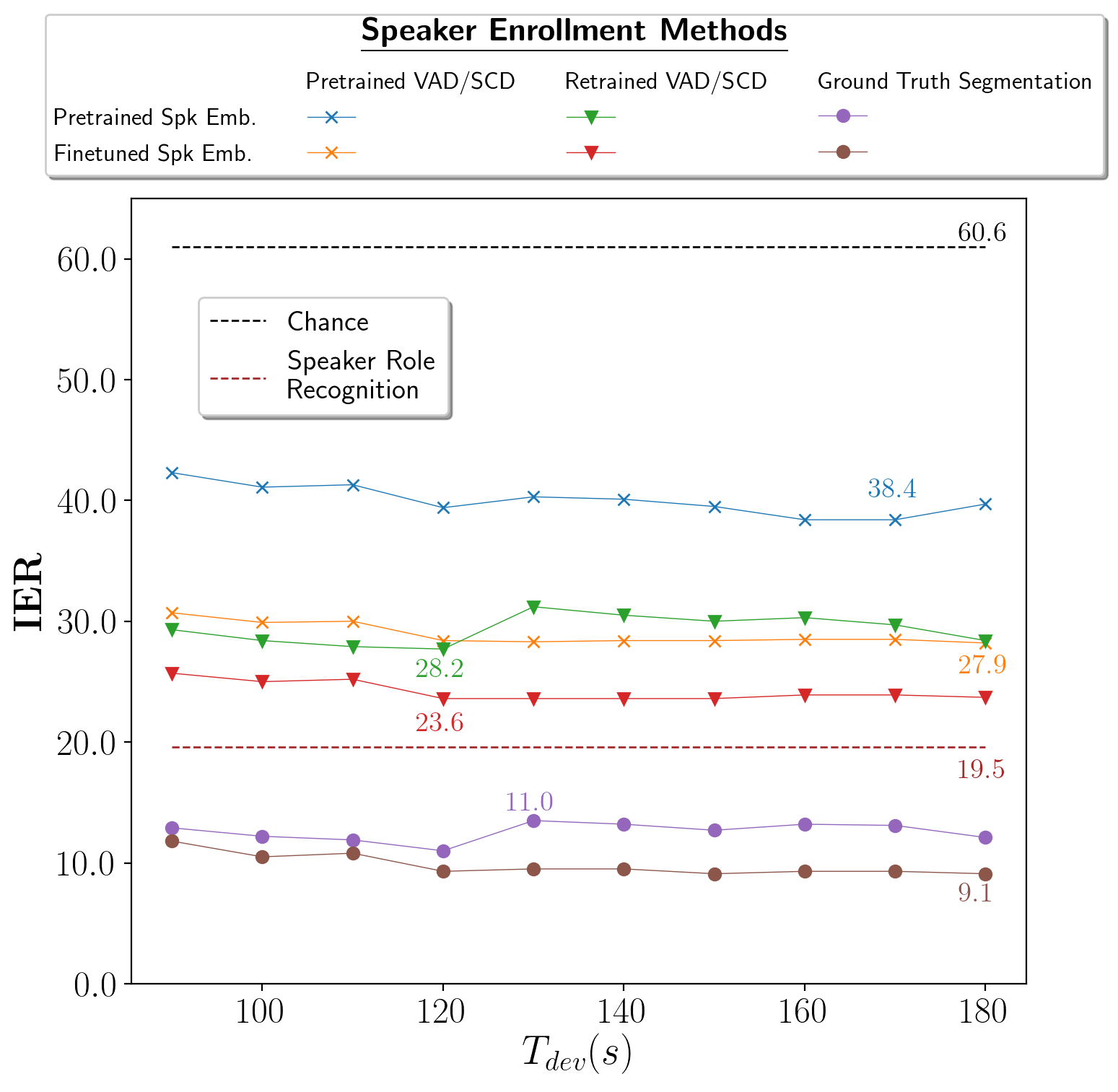}
\vspace{-1.em}
    \label{fig:results_t_dev}
\end{figure}

Then, we used the set of segments from the dev set $X_{dev}$ of the \textit{meta-dev} and \textit{meta-test} to build a template vector $m_j$ for each speaker $j$ in the interview $I$. $X_{dev}$ contain a set of segments $U_{\text{enrollment speaker } j}$ from each speaker $j$. The start of each segment $U_{\text{enrollment speaker } j}$ needs to be smaller than $T_{\text{dev}}$. We studied the effect of the size of the enrollment based on the parameter $T_{dev} \in (90s, 100s ,\dots, 170s, 180s)$ to build the template representation $m_j$.
We computed the average of the representations for each speaker $j$:
 \vspace{-0.5em}
\begin{equation}
            m_j = \frac{1}{|U_{\text{enrollment speaker } j}|}
        \sum_{U \in U_{\text{enrollment speaker } j}} f_{\theta}(U)
        \vspace{-1.em}
\end{equation}
\subsubsection{Identification}
 \vspace{-0.5em}
For the identification stage, we use the function $f_{\theta}$ and the different representation $m_j$ of the speakers from the enrollment stage. We used the following cosine distance $D$ to build a scoring function and compare each segment $\hat{U} \in \{\hat{U}_1, \dots \hat{U}_{N'_I}\}$ to each template $m_j$.
\begin{equation}
\vspace{-0.5em}
    D(\hat{U}, m_{j})= \frac{1}{2} \left(
    1 -
    \frac{f_{\theta}(\hat{U})^{\top} m_{j}}{ \left[\|f_{\theta}(\hat{U})\|\left\|m_{j}\right\|\right]} \right)
    \vspace{-0.5em}
\end{equation}

\begin{equation}
   \operatorname{argmin}_{j} D(\hat{U}, m_{j}) : \text{Selects Speaker } j
\end{equation}

In addition, we analyzed topline performance of the speaker embedding models when the Ground Truth Segmentation is provided.
Finally, we computed a chance baseline based on speaker Enrollment by randomly permutating all the cosine distances.

\vspace{-1.em}
\section{Results and discussions}
\vspace{-1.em}
\label{sec:results}
Figure \ref{fig:results_t_dev} shows results in term of IER for the different approaches. Both approaches greatly improved over chance. If we consider pipelines solving both segmentation and identification, our best performance is obtained using the Speaker Role Recognition approach with IER=19.5\% while the Speaker Enrollment Protocol obtained at best IER=23.6\% at $T_{dev}=120s$, with Retrained VAD/SCD models and Finetuned Speaker Embedding. Even though, the Speaker Enrollment protocol has per-speaker templates, it is not surpassing the Speaker Role Recognition approach. The topline with Ground Truth Segmentation (IER=9.1\%) indicated that Speaker Enrollment could benefit greatly from a better detection of speaker-homogeneous turns. Errors of Speaker Enrollment are accumulated through the steps and can not be recovered, while Speaker Role Recognition might take advantage of solving all steps together in its end-to-end approach.

\begin{figure}[!t]
    \centering
    \caption{Comparison of the Identification Error Rates per file on all the test sets $X_{test}$ of the Meta-test set $M_{test}$. The comparison is done for the Retrained/Finetuned approaches for the speaker enrollment protocol.}
    \includegraphics[width=0.4\textwidth]{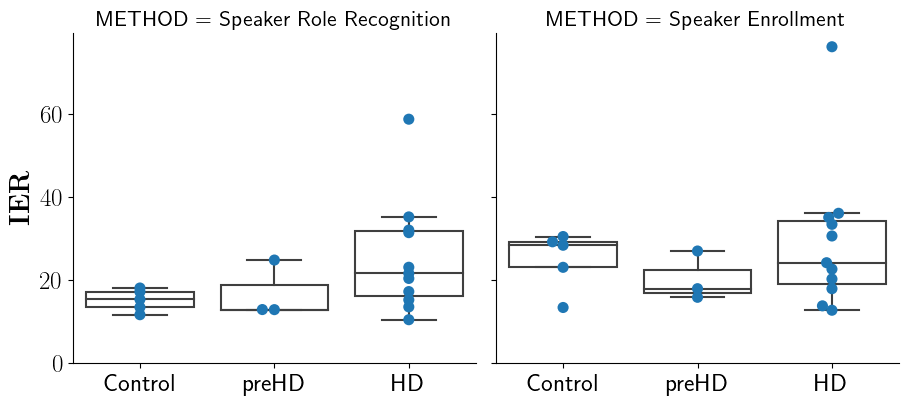}
\vspace{-1.5em}
    \label{fig:results_per_class}
\end{figure}

Increasing the size of the Template Enrollment $m_j$ for each speaker with $T_{dev}$ lead to light improvements to all Speaker Enrollment methods. The finetuning of the X-vector speaker embedding model with in-domain is especially crucial (ex: Based on retrained VAD/SCD the IER decreases from 28.2\% to 23.6\%).
An additional ablation study on the size of the meta-training set $M_{train}$ showed us that the IER goes from IER=19.5\% to IER=26.5\% for the Speaker Role Recognition model trained with only 10\% of $M_{train}$.
In Figure \ref{fig:results_per_class}, we showed the IER, for both approaches, along with the Interviewees' demographics. The IER is not collapsing for any of the group and for both methods, even for the Control which don't suffer from any speech disorder.

\vspace{-1.em}
\section{Conclusion and future work}
\vspace{-0.5em}
\label{sec:conclusions}
Detection and Identification of speaker turns are fundamental problems in speech processing, especially in healthcare applications. While works studying these problems in isolation has provided valuable insights, in this work, we showed that Speaker Role Recognition was the most suitable approach for Interviewees at different stages of Huntington's Disease. Lastly, we observed that both approaches have the potential to be used for clinical diagnosis. For future work, we plan to investigate the use of these methods to derive biomarkers automatically and compare them to classic approaches \cite{riad2020vocal} and extend it to recordings with a variable number of speakers.
\vspace{-1.5em}
\section{Acknowledgments}
\vspace{-0.5em}
We  are  very thankful  to  the  patients that participated in our study. We thank Katia Youssov, Laurent Cleret de Langavant, Marvin Lavechin, and the speech pathologists for the multiple helpful discussions and the evaluations of the patients.
This work is funded in part by the Agence Nationale pour la Recherche (ANR-17-EURE-0017 Frontcog, ANR-10-IDEX-0001-02 PSL*, ANR-19-P3IA-0001 PRAIRIE 3IA Institute) and Grants from Neuratris, from Facebook AI Research (Research Gift), Google (Faculty Research Award), Microsoft Research (Azure Credits and Grant), and Amazon Web Service (AWS Research Credits).



\newpage
\bibliographystyle{IEEEbib}
\small{
\bibliography{strings,refs}
}
\newpage
\section{Appendix}

Readers can find details of Inter-annotator agreements in \cite{seshat2020titeuxriad} where we measured the agreements to annotate speaker turns in these medical conversations between Interviewees.

\subsection{Ablation study}
\begin{table}[!h]
    \centering
    \caption{Speaker Role Recognition Ablation study: Identification Error Rates on the test set $X_{test}$ of the meta-test set $M_{test}$ as a function of the percentage of interview in the meta-train set $M_{train}$. MD stands for Missed detection, FA for False Alarm and Conf. for Confusion}
\begin{tabular}{|c|c|c|c|c|}

\hline
  Percentage of $M_{train}$  & MD & FA& Conf.& IER  \\
     \hline
    10\% &  $ 8.0$&  $ 14.5$&  $ 3.9$&  $ 26.5$ \\
        20\% &  $ 7.8$&  $ 12.4$&  $3.8$&  $24.0$\\
        50\%&  $ 7.5$&  $ 10.4$&  $2.5$&  $20.7$\\
        100\%&  $ 7.1$&  $ 10.2$&  $ 2.3$ &  $19.5$\\
    \hline
\end{tabular}
    \label{tab:results_m_train}
\end{table}

We ran an ablation study (Table \ref{tab:results_m_train}) for the Speaker Role Recognition to measure the amount of data necessary. Even though models are better than Chance, we found out that at least 50\% our dataset (28 Interviews) is necessary to outperform the Speaker Enrollment Protocol pipeline (20.7\% vs 23.6\%).

The details of the IER showed that the most important component is the False Alarm (FA), and the increase of size allowed to gain 4 points of FA.
\subsection{Clinical Relevance: Speech Markers}
\begin{figure}
    \centering
    \caption{Comparison of the Ratio of Silence between the Ground truth segmentation and the best Speaker role recognition pipeline. The comparison is done on the test sets $X_{test}$ of the Meta-test set $M_{test}$}
    \includegraphics[width=0.4\textwidth]{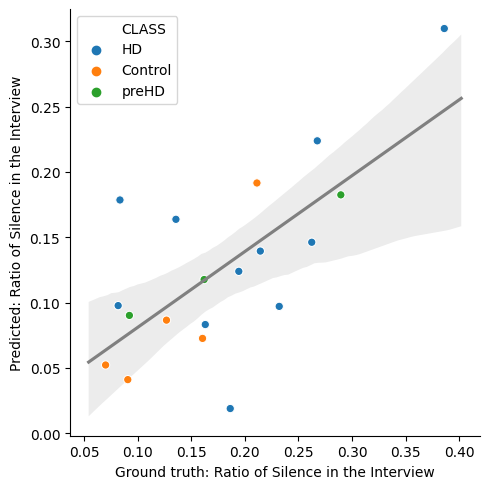}
    \label{fig:results_predicted_silence}
\end{figure}

\begin{figure}
    \centering
    \caption{Comparison of the Standard Deviations (SD) of the Duration of Utterances of Interviewees between the Ground truth segmentation and the best Speaker role recognition system. The comparison is done on the test sets $X_{test}$ of the Meta-test set $M_{test}$}
    \includegraphics[width=0.4\textwidth]{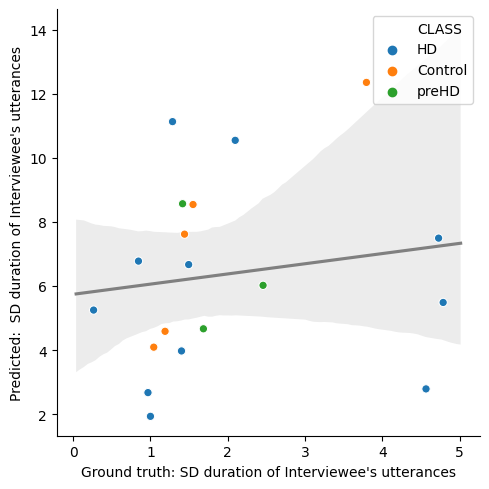}
    \label{fig:results_predicted_sd_utterance}
\end{figure}

In previous studies in Huntington's Disease \cite{vogel2012speech, perez2018classification}, the Ratio of Silence and Statistics on utterances and were informative to distinguish between classes of Individuals.  These speech markers can be extracted directly from the prediction of the Speaker Role Recognition outputs. We computed the Ratio of Silence and the Standard Deviation of Duration of Utterances on the test set of the Meta-test set $M_{test}$. This computation was done both from the Ground Truth Segmentation and the segmentation provided by the Speaker role recognition system, see Figure \ref{fig:results_predicted_silence} and Figure \ref{fig:results_predicted_sd_utterance}.

We found that the automatic system outputs behaved differently based on each speech marker. The Ratio of Silence was better predicted than the SD of Duration of Utterances. Some bias of the predictive system might not hurt the IER metric but hurt the reliability of some measures. Fortunately, the automatic system does not seem to have bias per class.

In future work, it would be valuable to take into account the speech markers in the loss and validation of the system.

\end{document}